\documentclass[conference]{IEEEtran}
\IEEEoverridecommandlockouts

\usepackage{cite}
\usepackage{amsmath,amssymb,amsfonts, mathtools}
\usepackage{algorithmic}
\usepackage{graphicx}
\usepackage{textcomp}
\usepackage{xcolor}
\usepackage{epstopdf}

\def\BibTeX{{\rm B\kern-.05em{\sc i\kern-.025em b}\kern-.08em
    T\kern-.1667em\lower.7ex\hbox{E}\kern-.125emX}} 
\graphicspath{{.}{../eps/}{./images/}{./images/result/}}
\begin{document}

\title{Quantifying Link Stability in Ad Hoc Wireless Networks Subject to Ornstein-Uhlenbeck Mobility
}

\author{\IEEEauthorblockN{Arta Cika\IEEEauthorrefmark{1}, Mihai-Alin Badiu\IEEEauthorrefmark{1}\IEEEauthorrefmark{2}, and Justin P. Coon\IEEEauthorrefmark{1}}
\IEEEauthorblockA{\IEEEauthorrefmark{1}
Department of Engineering Science\\
University of Oxford, Parks Road, Oxford, UK, OX1 3PJ\\
Email: \{arta.cika, mihai.badiu, and justin.coon\}@eng.ox.ac.uk}
\IEEEauthorblockA{\IEEEauthorrefmark{2}
Department of Electronic Systems\\
Aalborg University, Fredrik Bajers Vej 7, 9220 Aalborg {\O}st, Denmark}}

\maketitle

\begin{abstract}
The performance of mobile ad hoc networks in general and that of the routing algorithm, in particular, can be heavily affected by the intrinsic dynamic nature of the underlying topology. In this paper, we build a new analytical/numerical framework that characterizes nodes' mobility and the evolution of links between them. This formulation is based on a stationary Markov chain representation of link connectivity. The existence of a link between two nodes depends on their distance, which is governed by the mobility model. In our analysis, nodes move randomly according to an Ornstein-Uhlenbeck process using one tuning parameter to obtain different levels of randomness in the mobility pattern. Finally, we propose an entropy-rate-based metric that quantifies link uncertainty and evaluates its stability. Numerical results show that the proposed approach can accurately reflect the random mobility in the network and fully captures the link dynamics. It may thus be considered a valuable performance metric for the evaluation of the link stability and connectivity in these networks.
\end{abstract}


\section{Introduction}
Mobile ad hoc networks (MANET) consist of autonomous mobile nodes that can create a network in a decentralized manner, without the need for a fixed infrastructure~\cite{roy2010handbook, haas1999guest}. A link between two nodes exists if their received signal power is greater than a system-dependent threshold. As the nodes move away from each other, the link becomes inactive. In this environment, connections between nodes are established and broken intermittently causing the network topology to change over time. Because any node can behave as a router or a host, any changes in states of its links can affect every communication going through that node. Location uncertainty and link dynamics due to node mobility are thus a main factor impacting the performance of MANETs~\cite{chlamtac2003mobile}. Particularly, routing in these systems faces strong challenges due to the dynamically changing network topology~\cite{abolhasan2004review}. It is therefore imperative to take into account the random movements of nodes when designing and implementing these networks in any real-world application.

Many recent works used information theory tools to better understand the properties of complex networks. Studies in~\cite{coon2016topological, coon2017topological, cika2017effects, Bad18} used Shannon entropy to quantify the topological uncertainty of wireless networks embedded within a spatial domain, while in~\cite{coon2018conditional} different lower bounds on the Shannon entropy of random geometric graphs were derived by using the notion of conditional entropy. A way to identify critical nodes in a network using local vertex measures of entropy is presented in~\cite{tee2017vertex}. Rate-distortion theory was used in~\cite{wang2012cost} to characterize the minimum cost of tracking the motion state information of nodes in dynamic networks. In the field of MANET, a probabilistic method to assess the quality of the link in terms of link duration was proposed in~\cite{song2012link}, whereas~\cite{yawut2008mobility} studied three mobility metrics and evaluated their ability to predict the routing protocol performance. The authors in~\cite{an2002entropy} introduced an entropy-based model for evaluating route stability, and~\cite{tran2006link} proposed the entropy of the link change as a mobility metric. However, we believe entropy rate can represent a more accurate metric of link stability in dynamic networks, considering its ability to measure the uncertainty of the future state of the link given its current state.

In our previous work~\cite{cika2018entropy}, we used the entropy rate to analyze the topological uncertainty due to the variations in the propagation channel. In this paper, we present an information-theoretic framework for characterizing the uncertainty of the link connectivity due to node mobility and model the on-off transition as a stationary discrete-time Markov Chain. The crux of the problem lies in justifying the use of the Markov model; much of the present contribution is centered around this discussion. We consider a link between two nodes as active if the received instantaneous signal-to-noise ratio (SNR) is greater than a system-dependent threshold. In the absence of fading the only source of randomness in the state of the link is provided by the separation distance between nodes. In our model, nodes move randomly according to an Ornstein-Uhlenbeck (OU) process. This mean-reverting process is particularly suited for modeling node mobility in robotic swarms or D2D UAV networks subject to positional perturbations~\cite{kim2018massive}. The OU model represents a wide range of patterns with varying degrees of memory, including, as the two extreme cases, the random walk and the constant mobility model~\cite{camp2002survey}. For ad hoc environments where networks may change randomly and quickly, this mobility model is a good approach for the performance evaluation of these networks. 

Using this model, we formulate a mobility metric based on the notion of the entropy rate that can measure the link randomness and evaluate its stability. The proposed approach can be used as a source of information for routing protocols to quantify the minimum routing overhead necessary to maintain up-to-date topology information and to evaluate path stability to select the most stable route between two nodes.

The rest of the paper is organized as follows. Section~\ref{sec:system_model} provides the mobility model formulation, basic definitions and stationary condition. In section~\ref{sec:link_connectivity} we construct an analytical framework to model the existence of a link between two nodes as a Discrete Time Markov Chain (DTMC) with on and off states. We then introduce our link stability metric in section~\ref{sec:entropy_rate} and present the numerical results in section~\ref{sec:numerical}. Finally, the concluding remarks are discussed in section~\ref{sec:conclusions}.

\section{System Model}\label{sec:system_model}
Consider two arbitrary nodes (mobile wireless devices) moving randomly over a two-dimensional plane. Each device movement is assumed to be independent from the other. The locations of the nodes at time $t \geq 0$ are given by $Z_{1}(t)=\left(X_1(t),Y_1(t)\right)$ and $Z_{2}(t)=\left(X_2(t),Y_2(t)\right)$, respectively. We denote by $R(t)$ the Euclidean distance between two nodes, $R(t) = \|Z_2(t)-Z_1(t)\|$.

Next, let $L_{t_k}$ be a Bernoulli random variable that models the existence of the edge (link) between nodes in any arbitrary time step $t_k=t_0 + k\Delta t$, $k\in \mathbb{N}$. These time steps are finite and with equal duration defined by the constant $\Delta t>0$. A transmission from node $1$ to node $2$ at any time step $t_k$ is successful $\left(\text{the link is active}\right)$ if the SNR of the link, $\Gamma_{t_k}$, is greater than a certain threshold $\gamma_0$ determined by the communication hardware, as well as the modulation and coding scheme of the mobile ad hoc network. If we assume a SISO link between the two devices, then $\Gamma_{t_k}$ at any time step $t_k$ is given by
\begin{equation}
\Gamma_{t_k}=\psi R_{t_k}^{-\eta}|G_{t_k}|^2,
\end{equation}
where $\eta$ is the path loss exponent, $G_{t_k}$ is the fading channel gain with $E\left[|G_{t_k}|^2\right]=1$ and $\psi$ is a constant depending on different parameters such as transmit power, antenna properties, and wavelength. In this paper, we assume there is no fading affecting the link between nodes, i.e., $|G_{t_k}|^2=1$. It will become apparent that the omission of this detail does not hinder the development of important results. However, extensions to fading channels are left for consideration in the future. 
\subsection{Mobility Model}
In the following, we assume the initial positions of nodes  $1$ and $2$ to be $Z_1(0)=(0,0)$ and $Z_2(0)=(\beta,0)$, respectively. At time $t \geq 0$, their locations are given by  $Z_1(t)=(X_1(t),Y_1(t))$ and $Z_2(t)=(X_2(t),Y_2(t))$. We model the node displacements along the $x$ and $y$ coordinates, i.e. $\left\{X_1(t), X_2(t), Y_1(t), Y_2(t)\right\}$, by independent OU processes~\cite{uhlenbeck1930theory, doob1942brownian}. An OU process is defined as a continuous time stochastic process $\{S(t), t\geq0\}$ that satisfies the stochastic differential equation
\begin{equation}\label{eq:sde}
\mathrm{d}S(t)=\frac{1}{\tau}(\mu-S(t))\mathrm{d}t+\sqrt{D}\mathrm{d}W(t),
\end{equation}
where $\mu$ is the desired position, and $W(t)$ is the Wiener process. The parameters $\tau$ and $D$ are positive constants called the \textit{relaxation time} and the \textit{diffusion coefficient}, respectively; $\sqrt{D}$ controls the fluctuation in the position of the devices along each coordinate axis, and $1/\tau$ controls the rate of reversion of the device to the desired position (the initial position). Given the starting point $\{S_0, t=0\}$, the expectation and variance of the process are equal to\cite{gardiner2009stochastic}:
\begin{equation}\label{eq:ouprocess}
\begin{aligned}
\mathsf{E}\left[S(t)\right]&= \mu + (S_0-\mu)\exp\left[-t/\tau\right],\\
\mathsf{Var}\left[S(t)\right]&= \frac{D\tau}{2}\left(1-\exp\left[-2t/\tau\right] \right).
\end{aligned}
\end{equation}
Note that $\mathsf{Var}\left[S(t)\right]\rightarrow\frac{D\tau}{2}$ and $\mathsf{E}\left[S(t)\right]\rightarrow\mu$ as $t\rightarrow\infty$. The OU process is a Gaussian Markov process~\cite{gardiner2009stochastic}. If the initial condition of the process, $S_0$, is drawn according to the steady-state distribution, then the process is stationary. Another quantity of interest is the stationary correlation function of the OU process, which is obtained by allowing the system to approach its steady-state. It is given by~\cite{gardiner2009stochastic}
\begin{equation}\label{eq:autocorr}
\mathsf{E}\left\{\left[S(t)-m\right]\left[S(u)-m\right]\right\} = \frac{D\tau}{2}\exp\left[-\frac{|t-u|}{\tau}\right],
\end{equation}
where $m=\mathsf{E}\left[S(t)\right]$. In the steady-state, the random variables $S(t)$ and $S(u)$ are only significantly correlated if $|t-u|$ is equivalent to $\tau$, also known as the \textit{correlation time}~\cite{gardiner2009stochastic}.
Under the above described model, the random variables $X_1(t), Y_1(t), Y_2(t) \sim \mathcal{N}\left(0, \frac{D\tau}{2}\left(1-\exp\left[-2t/\tau\right] \right)\right)$ and $X_2(t) \sim \mathcal{N}\left(\beta, \frac{D\tau}{2}\left(1-\exp\left[-2t/\tau\right] \right)\right)$ are independent. On that account, we can write the separation distance between nodes at time $t$ as
\begin{equation}\label{eq:distance}
R(t)=\sqrt{ X^2(t)+Y^2(t)},
\end{equation}
where $X(t)=X_2(t)-X_1(t)\sim \mathcal{N}\left(\beta, D\tau\left(1-\exp\left[-2t/\tau\right]\right)\right)$ and $Y(t)=Y_2(t)-Y_1(t) \sim \mathcal{N}\left(0, D\tau\left(1-\exp\left[-2t/\tau\right]\right)\right)$ are independent random variables. By a simple transformation of random variables, it is easy to show that $R(t)\sim \mathrm{Rician}\left(\beta,\sqrt{g(t)}\right)$ for all $t$, and its probability density function is given by
\begin{equation}\label{eq:rician}
f_{R}(r;t)=\frac{r}{g(t)}\exp{\left[\frac{-\left(r^2+\beta^2\right)}{2g(t)}\right]}\mathrm{I}_0\left(\frac{\beta r}{g(t)}\right), 
\end{equation}
with $g(t)=D\tau\left(1-\exp\left[-2t/\tau\right]\right)$ and $\mathrm{I}_0$ being the modified Bessel function of the first kind with order zero.
\subsection{Discretization of the OU process}\label{subsec:discrete}
Instead of observing the locations of the mobile nodes continuously, we monitor them at regular time steps $t_k=t_0 + k\Delta t$, $k\in \mathbb{N}$ and $\Delta t > 0$. Thus, we can write the discrete version of the continuous time OU process $\{S(t), t\geq0\}$, valid for any positive value of $\Delta t$, as~\cite{gillespie1996exact}
\begin{multline}\label{eq:discrete_ou}
S_{t_k}=S_{t_k-1}\exp\left[-\Delta t / \tau\right]+\mu\left(1-\exp\left[-\Delta t/\tau\right]\right)\\
+\sqrt{\frac{D\tau\left(1-\exp\left[-2\Delta t/\tau\right]\right)}{2}}\epsilon_{t_k-1}, 
\end{multline}
where $\epsilon_{t_k}\sim\mathcal{N}\left(0,1\right)$ are independent and identically distributed Gaussian random variables. From~\eqref{eq:discrete_ou}, we observe the linear relationship between input and output in the form
\begin{equation}\label{eq:ar}
S_{t_k}=\phi S_{t_k-1}+\mu\left(1-\phi\right) +\sqrt{\frac{D\tau\left(1-\phi^2\right)}{2}}\epsilon_{t_k-1}\ k \in \mathbb{N},
\end{equation}
which corresponds to a first-order autoregressive process, $\mathrm{AR}(1)$, with parameter $\phi = \exp\left[-\Delta t/\tau\right]$. It is important to note that in our model $\phi\in \left(0,1\right)$ since $\Delta t > 0$ and $\tau > 0$. Therefore, the $\mathrm{AR}(1)$ process given in~\eqref{eq:ar} is stationary because the regression parameter $\phi$ satisfies the condition $|\phi|<1$, for every $\Delta t$~\cite{box2015time}.
\section{Markov Model of Link Connectivity}\label{sec:link_connectivity}
The purpose of this section is to obtain conditions under which a first-order Markov assumption can be applied. The key departure point is to treat the link state as a random process that exhibits random changes due to node mobility. In the absence of fading, at any time step $t_k$, the only source of randomness in the state of the link is provided by the separation distance $R_{t_k}=\sqrt{ X_{t_k}^2+Y_{t_k}^2}$, which is the discrete-time representation of~\eqref{eq:distance}\footnote{The discretization of the OU processes $X_1(t)$, $X_2(t)$, $Y_1(t)$, and $Y_2(t)$ is explained in section~\ref{subsec:discrete}.}. The random variable $L_{t_k}$ denotes the link state between nodes at any time step $t_k$, where $1 (0)$ defines whether the link exists (does not exist). Formally,
\begin{equation}
L_{t_k}=
\begin{cases}
1, & \mathrm{if}\ R_{t_k}\leq r_0, \\
0, & \mathrm{otherwise},
\end{cases}
\end{equation}
where $r_0 = (\psi/\gamma_0)^{\frac{1}{\eta}}$ is the typical connection range. This model is also known as the hard connection model of link connectivity~\cite{penrose2003random}. 

The first-order Markov assumption implies that the conditional distribution of $L_{t_k}$ depends only on $L_{t_k-1}$ and is independent from any other previous state. In the following, we will show that, under a set of constraints, the reduction in the uncertainty of $L_{t_k}$ due to knowledge of $L_{t_k-2}$, given the previous link state $L_{t_k-1}$, can be considered negligible. To make progress, we will need to evaluate the conditional probability $L_{t_k}$, which requires the knowledge of the joint probability distribution of $R_{t_k-2}$, $R_{t_k-1}$, and $R_{t_k}$. Now, consider the Gaussian vectors $\mathbf{X}=\left(X_{1}\ X_{2}\ X_{3}\right)^T$ and  $\mathbf{Y}=\left(Y_{1}\ Y_{2}\ Y_{3}\right)^T$. The $n$th component of the vector $\mathbf{R}$, $R_n=\sqrt{X_n^2+Y_n^2}$ where $n=1, 2, 3$, is a Rician random variable with probability density function given in~\eqref{eq:rician}. Furthermore, let $\mathbf{\Sigma}$ denote the covariance matrix of $\mathbf{X}$, and $\mathbf{W}=\mathbf{\Sigma}^{-1}$ its inverse with elements equal to $w_{uv}$, $1\leq u, v\leq 3$. Then the trivariate distribution of the Rician random variables $R_1$, $R_2$, and $R_3$ is given by~\cite{dharmawansa2008trivariate}
\begin{align}\label{eq:joint}\nonumber
&f_{R_1,R_2,R_3}\left(r_1,r_2,r_3\right)=\\\nonumber
&\frac{r_1r_2 r_3}{|\mathbf{\Sigma|}}\exp\left\{-\frac{1}{2}\left(\sum^3_{i=1}w_{ii}r^2_i+\beta^2w_4\right)\right\}\\
&\times\sum_{q=0}^\infty\sum_{p=-\infty}^\infty\varepsilon_k\left(-1\right)^{q+p}\mathrm{I}_q(w_3\beta r_3)\mathrm{I}_q(w_{32}\beta r_2r_3)\\\nonumber
&\times \mathrm{I}_p(w_1\beta r_1)\mathrm{I}_p(w_{12} r_1r_2)\mathrm{I}_{q+p}(w_{2} \beta r_2),
\end{align}
where $w_1=w_{11}+w_{12}$, $w_2=w_{22}+w_{23}+w_{12}$, $w_3=w_{33}+w_{23}$, $w_4=w_1+w_2+w_3$, $\mathrm{I}_n$ is the modified Bessel function of the first kind and order $n$, $|\mathbf{\Sigma}|$ is the determinant of the covariance matrix, and $\varepsilon_k$ is the Neumann factor $\left(\varepsilon_0=1, \varepsilon_n=2\ \mathrm{for}\ n=1,2, \dotsc\right).$ The joint probability distribution given in~\eqref{eq:joint} is valid only when $\mathbf{X}$ and $\mathbf{Y}$ have identical covariance matrix $\mathbf{\Sigma}$, and if $\mathbf{W}$ is a \textit{tridiagonal matrix} (i.e. $w_{13}=w_{31}=0$). Indeed, in our mobility formulation both these conditions are satisfied. The covariance matrices of $\mathbf{X}$ and $\mathbf{Y}$ are identical and equal to
\[
\mathbf{\Sigma}=D\tau\begin{bmatrix}
1 & \phi & \phi^2\\
\phi & 1 & \phi^2\\
\phi^2 & \phi & 1
\end{bmatrix},
\]
where $\phi = \exp\left[-\Delta t/\tau\right]$ is the regression parameter in~\eqref{eq:ar}. The matrix $\mathbf{\Sigma}$ has a Toeplitz structure, i.e. the correlation coefficients decay exponentially as the time shift between the elements of $\mathbf{\Sigma}$ increase. The inverse covariance matrix $\mathbf{W}$ has the tridiagonal property $\left(w_{13}=w_{31}=0\right)$, hence~\eqref{eq:joint} applies.
To better understand the coefficients of the covariance matrix $\mathbf{\Sigma}$, we refer to~\eqref{eq:ar}. Clearly, as $\phi\rightarrow 0$ or $\Delta t \rightarrow \infty $ while fixing $\tau$ the process described by~\eqref{eq:ar} represents a drifting random walk mobility pattern with mean $\mu$ and variance $ D\tau/2\left(1-\exp\left[-2t/\tau\right] \right)$. 
On the other hand, when $\phi\rightarrow 1$ or $\Delta t\rightarrow 0$ it degenerates into a constant mobility pattern with $S_{t_k}=S_0$ for all $k \in \mathbb{N}$. Therefore, the OU model represents a wide range of patterns with various degree of memory where $\phi$ can be seen as a tuning parameter to obtain different levels of random movement between these two extremities. 

For a fixed OU mobility model, the sampling interval $\Delta t$ controls the degree of memory in the stochastic process $\{L_{t_k}, k \in \mathbb{N}\}$. To assess the validity of the first-order Markov assumption we evaluate a mutual-information-based metric as a function of the sampling interval $\Delta t$. Given $L_{t_k-1}$, the importance of $L_{t_k-2}$ in providing information for $L_{t_k}$ can be measured by the ratio of the \textit{conditional mutual information} and the \textit{mutual information}~\cite{wang1996verifying}
\begin{equation}\label{eq:ratio}
\mathcal{R_{MI}} = \frac{I\left(L_{t_k};L_{t_k-2}|L_{t_k-1}\right)}{I\left(L_{t_k};L_{t_k-1},L_{t_k-2}\right)},
\end{equation}
for $t_k=t_0 + k\Delta t$, $k \in \mathbb{N}$. Without a closed-form solution for the joint probability distribution of $L_1$, $L_2$, and $L_3$, we numerically evaluate~\eqref{eq:ratio} for typical values of the tuning parameter $\phi \in \left(0 , 1\right)$. 

\begin{figure}[t]
\centering
\includegraphics[width = \columnwidth]{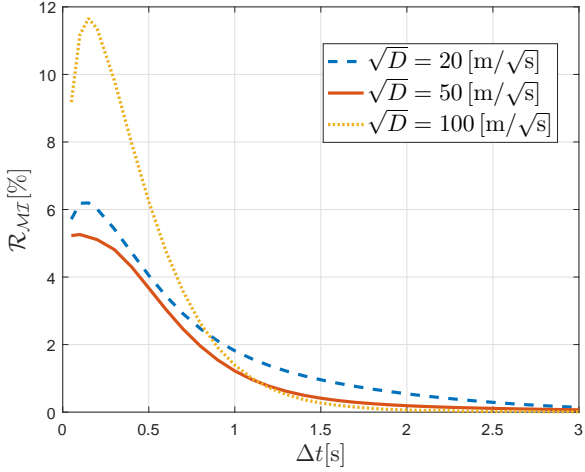}
\caption{Numerical evaluation for the mutual information ratio $\mathcal{R_{MI}}$ versus the sampling interval $\Delta t$; mobility parameters: $\tau=1\mathrm{s}$, $\sqrt{D} = 100 \mathrm{m}/\sqrt{\mathrm{s}}$, $\beta = 10\mathrm{m}$, and connection range $r_0=50\mathrm{m}$.}
\label{fig:mutual_info}
\end{figure}

In Fig.~\ref{fig:mutual_info} we plot the mutual information ratio $\mathcal{R_{MI}}$ versus the sampling interval $\Delta t$, for a given correlation time $\tau = 1s$ and for different values of the parameter $\sqrt{D}$. The diffusion coefficient $D$ in~\eqref{eq:sde} expresses the mean square distance traveled per unit of time~\cite{gardiner2009stochastic}. A few important things can be noted from the figure. First, as the sampling interval increases, the correlation between $L_{t_k-2}$ and $L_{t_k}$ gradually decreases leading to $\mathcal{R_{MI}} \rightarrow 0$. The opposite behavior is evident for small values of $\Delta t$, where the mutual information between $L_{t_k-2}$ and $L_{t_k}$ increases. However, for very small values of $\Delta t$ we notice a decrease in the value of the ratio $\mathcal{R_{MI}}$, particularly for low/high values of $\sqrt{D}$. The reason behind this behavior is that, fixing $\tau$, as $\sqrt{D}$ decreases or increases the node's movement becomes more or less restricted, respectively, causing less variations in the link state. This, in turn, reduces $I\left(L_{t_k};L_{t_k-2}|L_{t_k-1}\right)$, which measures the reduction in the uncertainty of $L_{t_k}$ due to knowledge of $L_{t_k-2}$ given $L_{t_k-1}$.
Second, the importance of $L_{t_k-2}$ given $L_{t_k-1}$ can be considered negligible $\left(\text{less than}\ 2\%\right)$ when $\Delta t \geq \tau $, as expected, given that $\tau$ indicates how strong the current node's location is correlated to its past ones. This behavior is verified for different values of the diffusion coefficient, as illustrated in Fig~\ref{fig:mutual_info}. Consequently, the first-order Markovian assumption is approximately verified when $\Delta t/ \tau$ satisfies the condition
\begin{equation}\label{eq:cond}
\frac{\Delta t}{\tau}\geq 1.
\end{equation}
Under this formalism, we can write
\begin{multline}
\mathsf{P}\left(L_{t_k}=a|L_{t_k-1}=b, L_{t_k-2}=c\right) \\
\simeq\mathsf{P}\left(L_{t_k}=a|L_{t_k-1}=b\right), 
\end{multline}
for all $a,b,c\in\{0,1\}$ and any time step $t_k$. To that end, we can approximate the stochastic process $\{L_{t_k}, k \in \mathbb{N}\}$ capturing the time evolution of the link between nodes as a stationary DTMC with transition probabilities
\begin{multline}\label{eq:cond_prob}
\mathsf{P}\left(L_{2}=a|L_{1}=b\right)=\\
\frac{\int_{r_3\in\mathbb{R}^+}\int_{r_2\in\mathcal{I}_a}\int_{r_1\in\mathcal{I}_b}f_{R_1,R_2,R3}\left(r_1,r_2,r_3\right)\mathrm{d}r_1\mathrm{d}r_2\mathrm{d}r_3}{\int_{{r\in\mathcal{I}_b}}f_R(r)\mathrm{d}r},
\end{multline}
and steady state probability
\begin{equation}\label{eq:steady_prob}
\mathsf{P}\left(L_{1}=b\right)=\int_{{r\in\mathcal{I}_b}}f_R(r)\mathrm{d}r,
\end{equation}
where the state variables $a ,b \in \{0,1\}$ determine the integration intervals $\mathcal{I}_a$ and $\mathcal{I}_b$, respectively. 
\begin{figure}[t]
\centering
\includegraphics[width = \columnwidth]{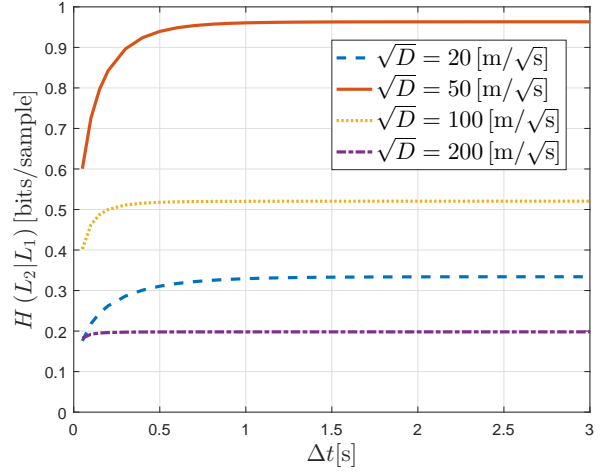}
\caption{Numerical evaluation for the entropy rate $H\left(L_2|L_1\right)$ versus the sampling interval $\Delta t$; mobility parameters: $\tau = 1\mathrm{s}$, $\beta = 10\mathrm{m}$, connection range $r_0=50\mathrm{m}$.}
\label{fig:EntropyRateVsDt}
\end{figure}
\section{Entropy Rate as a Link Stability Metric}\label{sec:entropy_rate}
In this section, we introduce a metric based on the information theoretic notion of entropy rate to evaluate the link stability in mobile ad hoc networks. The random movements of the nodes produce a sequence of on-off links leading to frequently changing network connectivity. In our analysis, the link state evolution $\{L_{t_k}, k \in \mathbb{N}\}$ is modeled as a stationary Markov chain, and its entropy rate is equal to the transition entropy~\cite{cover2012elements}
\begin{multline}
H\left(L_2|L_1\right)=-\sum_{b\in\{0,1\}} \mathsf{P}\left(L_1=b\right)\\
\times \sum_{a\in\{0,1\}}\mathsf{P}\left(L_2=a|L_1=b\right)\log_2\mathsf{P}\left(L_2=a|L_1=b\right).
\end{multline}
We can interpret the entropy rate as a measure of the uncertainty of the future state of the link given its past states. It is important to note that the OU model is not strictly stationary since its statistics in~\eqref{eq:ouprocess} depend on time $t$. However, for the entropy rate, the transient behavior is suppressed when we take the limit $H\left(L_2|L_1\right)=\lim_{k\rightarrow\infty}H\left(L_{t_k}|L_{t_k-1},\dotsc,L_1\right)$, and $t_k=t_0 + k\Delta t$, $k\in \mathbb{N}$.

The transition probability, $\mathsf{P}\left(L_2=a|L_1=b\right)$, reflects the random mobility in the network and fully captures the link dynamics. The distance $R_k$ changes due to the movement of nodes influencing the value of random variable $L_k$, and therefore of $H\left(L_2|L_1\right)$. The entropy rate quantifies how quickly the link state is varying with time. So, a high entropy rate indicates that the link is frequently changing over time. Thus, the metric $H\left(L_2|L_1\right)$ can be considered a good approach to characterize the uncertainty due to the random mobility. 

From an information theoretic perspective, the entropy rate measures the average minimum description length of the stochastic process capturing the link dynamics. Hence, it can quantify the minimum routing overhead necessary to maintain up-to-date topology information across a mobile ad hoc network. It can also be used as a source of information for routing protocols to quantify path stability and select the most stable route between two nodes. A broader analysis of the beneficial effects of the entropy rate on routing algorithms is left for consideration in the future.
\begin{figure}[t]
\centering
\includegraphics[width = \columnwidth]{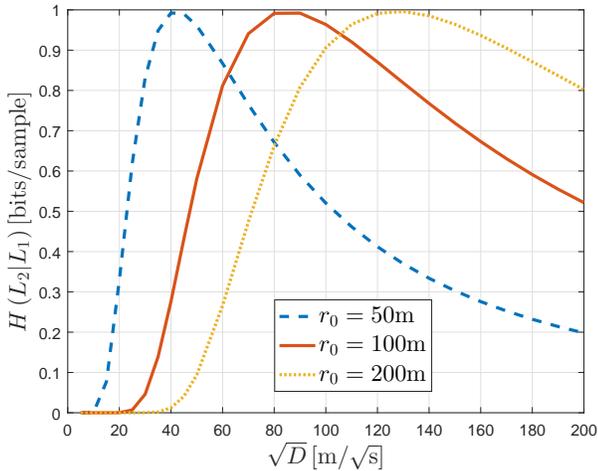}
\caption{Numerical evaluation for the entropy rate $H\left(L_2|L_1\right)$ versus the square root of the diffusion coefficient $\sqrt{D}$; mobility parameters: $\tau = 1\mathrm{s}$, $\Delta t =1\mathrm{s}$, and $\beta = 10\mathrm{m}$.}
\label{fig:EntropyRateVsSD}
\end{figure}
\section{Numerical Results and Discussions}\label{sec:numerical}
There are three important parameters in our model that affect the movement of nodes, and therefore their separation distance:  $\Delta t, \tau$, and $\sqrt{D}$. It is of fundamental interest to understand the impact of these parameters on the link stability metric. Without a closed-form solution for the joint probability density of $L_1$, $L_2$, and $L_3$, we calculate the entropy rate by numerically evaluating~\eqref{eq:cond_prob} and~\eqref{eq:steady_prob}. This evaluation is performed for typical values of the tuning parameter $\phi \in \left(0 , 1\right)$ which determines the intervals of $\Delta t$ and $\tau$. 

In Fig.~\ref{fig:EntropyRateVsDt} we analyze the behavior of the entropy rate $H\left(L_2|L_1\right)$ versus the sampling interval $\Delta t$, for a given correlation time $\tau$. As the sampling interval increases the correlation between $L_2$ and $L_1$ gradually decreases. Consequently, the transition entropy asymptotically converges to its maximum value $H\left(L_2\right)$. Here, we also explore the impact of the parameter $\sqrt{D}$ on the link stability metric. It is clear from the figure that when $\sqrt{D}$, which controls the fluctuation in the position of the devices, is very close in value to the connection range there is maximum uncertainty in the link state. This makes perfect sense, as in this scenario the separation distance oscillates around $r_0$. In the hard connection model nodes are connected whenever they lie within some critical distance of each other. Therefore, the link state between two nodes lying at the border of each other's radio range is characterized by maximum uncertainty. Instead, when $\sqrt{D}$ increases or decreases with respect to $r_0$, the link uncertainty decreases and for very high values, $\sqrt{D}=200\left[\mathrm{m}/\sqrt{\mathrm{s}}\right]$, the transition entropy is the same for any $\Delta t$.
\begin{figure}[t]
\centering
\includegraphics[width = \columnwidth]{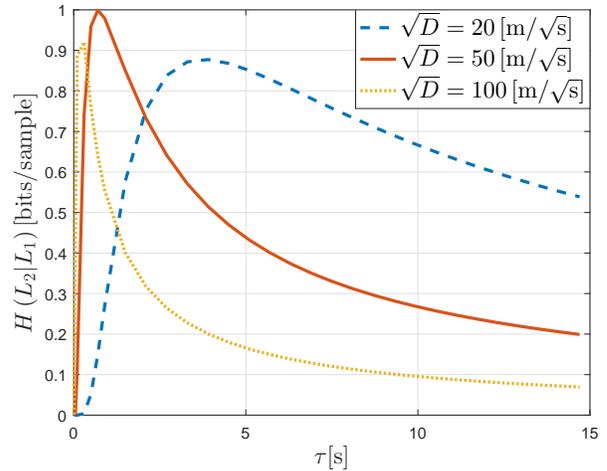}
\caption{Numerical evaluation for the entropy rate $H\left(L_2|L_1\right)$ versus the correlation time $\tau$; mobility parameters: $\Delta t = 1\mathrm{s}$, $\beta = 10\mathrm{m}$, and connection range = $50\mathrm{m}$.}
\label{fig:EntropyRateVsTau}
\end{figure}

The impact of the square root of the mean square distance traveled per unit of time on the entropy rate is shown in Fig.~\ref{fig:EntropyRateVsSD}. Note that the transition entropy approaches its maximum value $\left(1\left[\mathrm{bit}/\mathrm{sample}\right]\right)$ as $\sqrt{D}\approx r_0$. In contrast, the transition entropy decreases as $\sqrt{D}\rightarrow 0$ and $\sqrt{D}\rightarrow \infty$. This behavior verifies the observations made previously. Intuitively, one can recognize that, fixing $\tau$, a node's instantaneous displacement becomes more unrestricted with the increase of $\sqrt{D}$. This signifies that nodes will be less likely to be in the radio range of each other, reducing the uncertainty of the link state. In the same fashion, for small values of $\sqrt{D}$ the movement of nodes can be very restricted. Two nodes initially connected/disconnected will continue to remain so, introducing memory in the system. Hence, the uncertainty of the future state of the link given its current state decreases. 

Finally, in Fig.~\ref{fig:EntropyRateVsTau}  we investigate the effect of the relaxation time on the entropy rate. Fixing $\Delta t$, as $\tau$ increases the randomness in the link state decreases. This is straightforward from the fact that an increase in $\tau$ suggests that the node's current and past displacements are becoming more correlated, decreasing the uncertainty of the link future state given its current one. On the contrary, as $\tau$ decreases the opposite behavior is observed, where the uncertainty of the future state of the link given its current state increases. However, for small values of $\tau$ it is shown in Fig.~\ref{fig:EntropyRateVsTau} that with the decrease of $\tau$ the entropy rate is actually decreasing. The reason behind this initial decay on the entropy rate, which may seem counterintuitive, is related to the variance of node's displacement along the $x$-axis and $y$-axis. From~\eqref{eq:ar} we have
\begin{equation}
\mathrm{var}\left[S_k\right]=\frac{D\tau}{2}.
\end{equation}
The variance of an OU process is controlled by $\tau$ and $D$. For small values of $\tau$ while fixing $D$ the node's displacement can become very restricted. Therefore, their separation distance will vary less, reducing the uncertainty in the link state. 
\section{Conclusions}\label{sec:conclusions}

In this work, we developed a framework based on the entropy rate measure for evaluating link stability in mobile ad hoc networks. We started our analysis by modeling the existence of an edge between two nodes as a stationary Markov chain whose source of randomness is the separation distance between the nodes. Using this model, we formulated an entropy-rate-based metric that evaluates the link stability. The presented metric takes full advantage of the correlation between the link current and its future state. The motivations behind this work arose from the intrinsic location uncertainty of MANETs and the ability of the entropy rate to capture this randomness. We applied our calculations to nodes experiencing an OU mobility model and analyzed the impact of the mobility parameters on the link stability. Finally, through numerical results, we demonstrated that the proposed scheme thoroughly captures the link dynamics and is able to accurately reflect changes in the state of a wireless connection. 

Future work can be focused on analyzing the beneficial effects of the entropy rate on routing algorithms and on the performance of MANETs in general. The derived framework can be used to design new algorithms that adjust their operating mode based on link stability. The proposed concepts and approaches can be extended to different mobility models, fading channels and networks.
\section*{Acknowledgment}
The authors wish to acknowledge the support of Moogsoft and EPSRC under grant number EP/N002350/1 (``Spatially Embedded Networks''). The authors would like to thank Phil Tee from Moogsoft for his comments and feedback. 
\IEEEtriggeratref{3}
\bibliographystyle{ieeetr}
\bibliography{IEEEabrv,biblio}

\begin{thebibliography}{10}

\bibitem{roy2010handbook}
R.~R. Roy, {\em Handbook of mobile ad hoc networks for mobility models}.
\newblock Springer Science \& Business Media, 2010.

\bibitem{haas1999guest}
Z.~J. Haas, M.~Gerla, D.~B. Johnson, C.~E. Perkins, M.~B. Pursley,
  M.~Steenstrup, C.-K. Toh, and J.~F. Hayes, ``Guest editorial wireless ad hoc
  networks,'' {\em IEEE Journal on Selected Areas in Communications}, vol.~17,
  no.~8, pp.~1329--1332, 1999.

\bibitem{chlamtac2003mobile}
I.~Chlamtac, M.~Conti, and J.~J.-N. Liu, ``Mobile ad hoc networking:
  imperatives and challenges,'' {\em Ad hoc networks}, vol.~1, no.~1,
  pp.~13--64, 2003.

\bibitem{abolhasan2004review}
M.~Abolhasan, T.~Wysocki, and E.~Dutkiewicz, ``A review of routing protocols
  for mobile ad hoc networks,'' {\em Ad hoc networks}, vol.~2, no.~1,
  pp.~1--22, 2004.

\bibitem{coon2016topological}
J.~P. Coon, ``Topological uncertainty in wireless networks,'' in {\em Global
  Communications Conference (GLOBECOM), 2016 IEEE}, pp.~1--6, IEEE, 2016.

\bibitem{coon2017topological}
J.~P. Coon and P.~J. Smith, ``Topological entropy in wireless networks subject
  to composite fading,'' in {\em Communications (ICC), 2017 IEEE International
  Conference on}, pp.~1--7, IEEE, 2017.

\bibitem{cika2017effects}
A.~Cika, J.~P. Coon, and S.~Kim, ``Effects of directivity on wireless network
  complexity,'' in {\em Modeling and Optimization in Mobile, Ad Hoc, and
  Wireless Networks (WiOpt), 2017 15th International Symposium on}, pp.~1--7,
  IEEE, 2017.

\bibitem{Bad18}
M.~Badiu and J.~P. Coon, ``On the distribution of random geometric graphs,'' in
  {\em 2018 IEEE International Symposium on Information Theory (ISIT)},
  pp.~2137--2141, June 2018.

\bibitem{coon2018conditional}
J.~P. Coon, M.~A. Badiu, and D.~G{\"u}nd{\"u}z, ``On the conditional entropy of
  wireless networks,'' in {\em 2018 International Workshop on Spatial
  Stochastic Models for Wireless Networks (spaswin)-Invited Paper}, 2018.

\bibitem{tee2017vertex}
P.~Tee, G.~Parisis, and I.~Wakeman, ``Vertex entropy as a critical node measure
  in network monitoring,'' {\em IEEE Transactions on Network and Service
  Management}, vol.~14, no.~3, pp.~646--660, 2017.

\bibitem{wang2012cost}
D.~Wang and A.~A. Abouzeid, ``On the cost of knowledge of mobility in dynamic
  networks: An information-theoretic approach,'' {\em IEEE Transactions on
  Mobile Computing}, vol.~11, no.~6, pp.~995--1006, 2012.

\bibitem{song2012link}
Q.~Song, Z.~Ning, S.~Wang, and A.~Jamalipour, ``Link stability estimation based
  on link connectivity changes in mobile ad-hoc networks,'' {\em Journal of
  Network and Computer Applications}, vol.~35, no.~6, pp.~2051--2058, 2012.

\bibitem{yawut2008mobility}
C.~Yawut, B.~Paillassa, and R.~Dhaou, ``Mobility metrics evaluation for
  self-adaptive protocols.,'' {\em JNW}, vol.~3, no.~1, pp.~53--64, 2008.

\bibitem{an2002entropy}
B.~An and S.~Papavassiliou, ``An entropy-based model for supporting and
  evaluating route stability in mobile ad hoc wireless networks,'' {\em IEEE
  Communications Letters}, vol.~6, no.~8, pp.~328--330, 2002.

\bibitem{tran2006link}
Q.-M. Tran, A.~Dadej, and S.~Perreau, ``Link change and generalized mobility
  metric for mobile ad-hoc networks,'' in {\em Military Communications
  Conference, 2006. MILCOM 2006. IEEE}, pp.~1--7, IEEE, 2006.

\bibitem{cika2018entropy}
A.~Cika, M.-A. Badiu, J.~P. Coon, and S.~E. Tajbakhsh, ``Entropy rate of
  time-varying wireless networks,'' in {\em Ieee Global Communications
  Conference (globecom)}, in press, 2018.

\bibitem{kim2018massive}
H.~Kim, J.~Park, M.~Bennis, and S.-L. Kim, ``Massive uav-to-ground
  communication and its stable movement control: A mean-field approach,'' {\em
  arXiv preprint arXiv:1803.03285}, 2018.

\bibitem{camp2002survey}
T.~Camp, J.~Boleng, and V.~Davies, ``A survey of mobility models for ad hoc
  network research,'' {\em Wireless communications and mobile computing},
  vol.~2, no.~5, pp.~483--502, 2002.

\bibitem{uhlenbeck1930theory}
G.~E. Uhlenbeck and L.~S. Ornstein, ``On the theory of the brownian motion,''
  {\em Physical review}, vol.~36, no.~5, p.~823, 1930.

\bibitem{doob1942brownian}
J.~L. Doob, ``The brownian movement and stochastic equations,'' {\em Annals of
  Mathematics}, pp.~351--369, 1942.

\bibitem{gardiner2009stochastic}
C.~Gardiner, {\em Stochastic methods}, vol.~4.
\newblock springer Berlin, 2009.

\bibitem{gillespie1996exact}
D.~T. Gillespie, ``Exact numerical simulation of the ornstein-uhlenbeck process
  and its integral,'' {\em Physical review E}, vol.~54, no.~2, p.~2084, 1996.

\bibitem{box2015time}
G.~E. Box, G.~M. Jenkins, G.~C. Reinsel, and G.~M. Ljung, {\em Time series
  analysis: forecasting and control}.
\newblock John Wiley \& Sons, 2015.

\bibitem{penrose2003random}
M.~Penrose, {\em Random geometric graphs}.
\newblock No.~5, Oxford University Press, 2003.

\bibitem{cover2012elements}
T.~M. Cover and J.~A. Thomas, {\em Elements of information theory}.
\newblock John Wiley \& Sons, 2012.

\bibitem{dharmawansa2008trivariate}
P.~Dharmawansa, N.~Rajatheva, and C.~Tellambura, ``On the trivariate rician
  distribution,'' {\em IEEE Transactions on communications}, vol.~56, no.~12,
  pp.~1993--1997, 2008.

\bibitem{wang1996verifying}
H.~S. Wang and P.-C. Chang, ``On verifying the first-order markovian assumption
  for a rayleigh fading channel model,'' {\em IEEE Transactions on Vehicular
  Technology}, vol.~45, no.~2, pp.~353--357, 1996.

\end{thebibliography}
\end{document}